\documentclass[journal,comsoc]{IEEEtran}
\IEEEoverridecommandlockouts
\usepackage{cite}
\usepackage{hyperref} 
\usepackage{amsmath,amssymb,amsfonts}
\usepackage{algorithmic}
\usepackage{graphicx}
\usepackage{textcomp}
\usepackage{xcolor}
\usepackage{algorithm}
\usepackage{algorithmic}
\usepackage{tabularx}
\usepackage{tabu}  
\usepackage{float}
\usepackage{epstopdf}
\usepackage{cite} 
\usepackage{subfigure}
\usepackage{setspace}
\usepackage{multirow}
\usepackage{booktabs} 
\usepackage{makecell}
\usepackage{authblk}
\def\BibTeX{{\rm B\kern-.05em{\sc i\kern-.025em b}\kern-.08em
    T\kern-.1667em\lower.7ex\hbox{E}\kern-.125emX}}

\begin{document}
\title{Interplay between RIS and AI in Wireless Communications: Fundamentals, Architectures, Applications, and Open Research Problems}

\author[1]{Jinghe Wang}
\author[1]{Wankai Tang}
\author[1]{Yu Han}
\author[1]{Shi Jin}
\author[1]{Xiao Li}
\author[2]{\\ Chao-Kai Wen}
\author[3]{Qiang Cheng}
\author[3]{Tie Jun Cui}

\affil[1]{National Mobile Communications Research Laboratory, Southeast University, Nanjing, China, \authorcr (email: \{wangjh, tangwk, hanyu, jinshi, li\_xiao\}@seu.edu.cn)}
\affil[2]{Institute of Communications Engineering, National Sun Yat-sen University, Kaohsiung, Taiwan, \authorcr (email: chaokai.wen@mail.nsysu.edu.tw)}
\affil[3]{State Key Laboratory of Millimeter Waves, Southeast University, Nanjing, China, \authorcr (email: \{qiangcheng, tjcui\}@seu.edu.cn)}

\renewcommand*{\Affilfont}{\small\it} 

\maketitle
\begin{abstract}
Future wireless communication networks are expected to fulfill the unprecedented performance requirements to support our highly digitized and globally data-driven society. Various technological challenges must be overcome to achieve our goal. Among many potential technologies, reconfigurable intelligent surface (RIS) and artificial intelligence (AI) have attracted extensive attention, thereby leading to a proliferation of studies for utilizing them in wireless communication systems. The RIS-based wireless communication frameworks and AI-enabled technologies, two of the promising technologies for the sixth-generation networks, interact and promote with each other, striving to collaboratively create a controllable, intelligent, reconfigurable, and programmable wireless propagation environment. 
This paper explores the road to implementing the combination of  RIS and AI; specifically, integrating AI-enabled technologies into RIS-based frameworks for maximizing the practicality of RIS to facilitate the realization of smart radio propagation environments, elaborated from shallow to deep insights. We begin with the basic concept and fundamental characteristics of RIS, followed by the overview of the research status of RIS. Then, we analyze the inevitable trend of RIS to be combined with AI. In particular, we focus on recent research about RIS-based architectures embedded with AI, elucidating from the intelligent structures and systems of metamaterials to the AI-embedded RIS-assisted wireless communication systems. Finally, the challenges and potential of the topic are discussed.
\end{abstract}

\begin{IEEEkeywords}
6G, metamaterials, reconfigurable intelligent surface, artificial intelligence, smart radio propagation environment
\end{IEEEkeywords}

\section{Introduction}

\IEEEPARstart{W}{ith} the global standardization and commercialization of the fifth-generation (5G) networks by 2020, the communication academia and industry are devoted to shaping the next-generation communication system, which is the sixth-generation (6G). The 6G networks will be able to fulfill unprecedented performance requirements to make all through hyper-connectivity involving humans and machines possible and our society will be highly digitized and globally data-driven by providing the ubiquitous and reliable wireless connectivity  \cite{samsung}. 
On the one hand, typical frameworks of 5G, such as enhanced mobile broadband (eMBB), ultra-reliable and low latency communications (uRLLC), and massive machine-type communications (mMTC) are expected to obtain continuous improvement towards next generation \cite{andrews2014will,Simsek20165G,20175G}. On the other hand, 6G yearns for ground breaking paradigm shifts. Advances in communications, high-resolution imaging and sensing, accurate identification and positioning, mobile robots and drone techniques lead to the emergence of brand-new services and applications; examples of these applications include, truly immersive multisensory extended reality (XR) services, connected robotics and autonomous systems, high-fidelity mobile hologram, wireless brain-computer interactions, blockchain, and distributed ledger technologies, which extremely make our daily life smoother and safer and significantly improve the enterprise and individual efficiency \cite{latva2020key,zhang20196g,saad2019vision,2020A}. 

To realize these exciting applications, the performance requirements of these services include a peak data rate of 1 Tbps and over-the-air latency of less than 100 $\mu \mathrm{s}$, raising the typical peak data rate by orders-of-magnitude compared with 5G and one-tenth the latency of 5G and, supporting broadband connectivity at railway speeds up to 1000 km/h \cite{rajatheva2020white}. Considering the challenges for guaranteeing these performance requirements, the way the data are collected, processed, transmitted, and received over the 6G wireless system is expected to be innovated or even redefined. 
Following this trend, rather than merely adapting to the communication environment for acquiring better system performance by elaborately designing the transmit or receive modules of the systems through large-scale multiple-input multiple-output (MIMO) \cite{lu2014overview}, multiplexing and diversity, beamforming, and precoding technologies \cite{tse2005fundamentals,zheng2003diversity,ahmed2018survey}, which lack the adjustment of wireless propagation environment in current 5G wireless networks, 6G networks may have the vision of shaping the radio propagation environment towards their liking \cite{9140329}.

With the revolution in electromagnetic (EM) metamaterials \cite{cui2014coding,cui2017information,zhang2018space,bao2018design,zhao2019programmable}, the reconfigurable intelligent surface (RIS) has received considerable attention in recent years due to its unique characteristics of EM wave regulation, therefore rapidly becoming a key instrument in realizing the intelligence of the propagation environment. Based on RISs, 6G networks can tailor the propagation paths of the signal. Such a transformative wireless concept of tailored radio propagation achieves significant developmental potential and application value. The RISs can not only break through the half-wavelength limitation of antenna spacing in wireless communication systems but also gain the advantages of cost and energy efficiencies, according to which, RISs can be spread over the whole city to promptly establish smooth communication links, embracing everything in internet-of-things (IoT). In addition to providing the supplementary links and achieving the propagation environment reconfiguration, RISs can recycle EM waves, which can effectively reduce the power consumption for uninterrupted data transmission. Smart radio propagation environments based on RISs have the potential to provide ubiquitous and uninterrupted wireless connections in the most energy-efficient way. 

\begin{figure}[t]   
	\centering  
	\includegraphics[height=6.3cm,width=9cm]{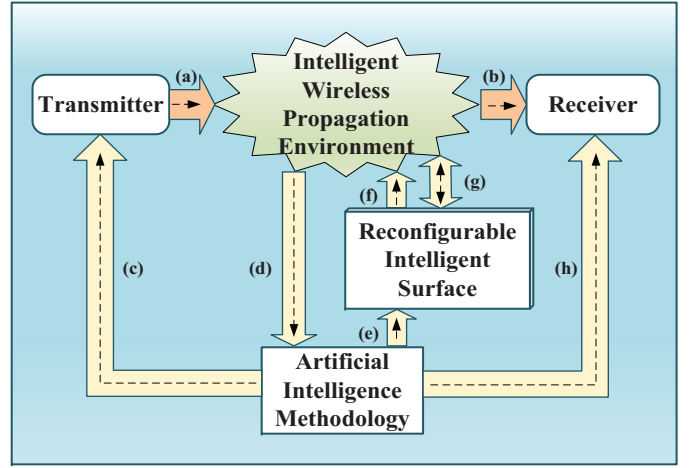}  
    \vspace{-0.3cm}
	\caption{Paradigm of configuring an intelligent wireless propagation environment (IWPE) of 6G by RIS and AI. (a) and (b): Data flow. (c), (d) and (h): AI learns the environmental information to design the transmitter and receiver. (e) and (f): AI configures the IWPE through RIS. (g): RISs interact with IWPE. }  
	\label{img1}  
	\vspace{-0.3cm}
\end{figure}

Besides RISs, artificial intelligence (AI)  \cite{russel2013artificial,schmidhuber2015deep,jordan2015machine} has received considerable attention and widespread recognition for assisting wireless communication systems as early as 5G. Moreover, AI has the potential to satisfy the performance of future wireless communication networks and play a major role in the link-level and system-level solutions of 6G. The RIS-based wireless communication frameworks and AI-enabled technologies, two of the promising technologies for future wireless networks, interact and promote with each other, striving to create a controllable, intelligent, reconfigurable, and programmable wireless propagation environment. The paradigm of configuring IWPE is shown in Fig.~\ref{img1}. Therefore, combining RIS and AI techniques is imperative. On this basis, we explore, analyze, and overview the road to implementing RIS architectures, which are integrated with AI in 6G wireless networks. 

The rest of this paper is organized as follows. The basic concepts, fundamental characteristics, and research status of RISs are presented in Section II. The inevitable trend of RIS combined with AI is also thoroughly elucidated. A synopsis of how AI can be embedded with RISs frameworks is provided in Sections III to V. The intelligent structures of metamaterials are introduced in Section III. The intelligent systems of metamaterials are expounded in Section IV. The AI-embedded RIS-assisted wireless communication systems are elaborately discussed in Section V. Then, the challenges and potentials related to the topic are provided and discussed in Section VI. Finally, the conclusions are summarized in Section VII.

\section{Reconfigurable Intelligent Surface}

\begin{figure*}[t]   
	\centering  
	\includegraphics[height=5cm,width=18cm]{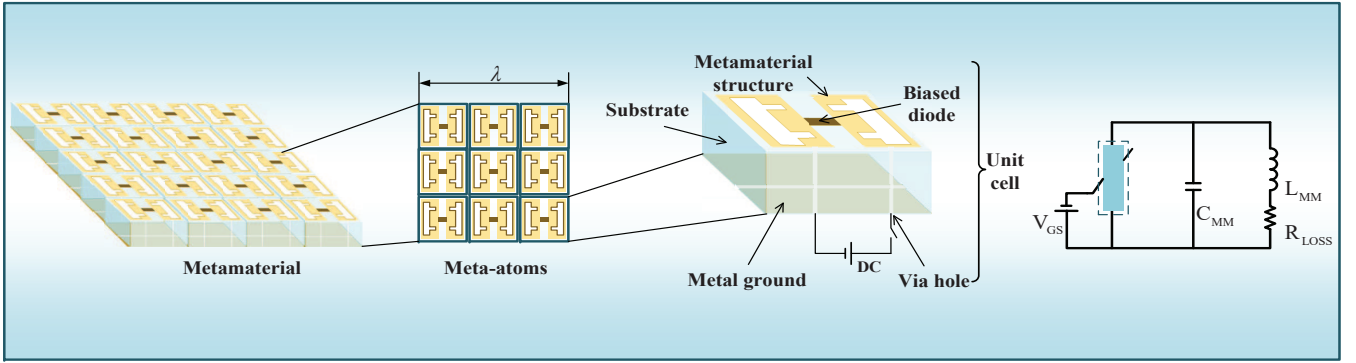} \caption{Physical structure of the metamaterial and its unit cell.}  
	\label{img2}  
	\vspace{-0.2cm}
\end{figure*}
  
The future wireless networks tend to use high frequency spectrum. Accordingly, the wireless propagation conditions become tougher and more challenging because extreme penetration losses and fewer scatterers lead to the channel sparsity and rank deficiency and shortage of the available links between the transmitter and the user. Meanwhile, the beam design of high-frequency antenna arrays also becomes difficult. Under such conditions, RISs can be deployed and utilized to provide supplementary links to improve the propagation environments in a cost efficient and energy efficient way, which bring new degrees of freedom for the system performance enhancement \cite{2019Smart,huang2019reconfigurable,wu2019towards,chen2019intelligent,basar2019wireless,2020MIMO,del2019optimally}. In this section, we start from the fundamental knowledge and characteristics of metamaterials, metasurfaces, and RISs  and then present the research status of RIS utilized in wireless communications. Furthermore, the inevitable trend of RIS to be combined with AI is also thoroughly explicated. 

\subsection{Metamaterial, metasurface, and RIS}
Metamaterial, also called special media or new artificial EM media, is not the material that can be naturally synthesized in nature. In Fig.~\ref{img2}, the physical structure of the metamaterial is created by imitating the lattice arrangement of atoms or molecules in nature and periodically arranging the artificial units of subwavelengths to interact with external EM fields through resonant systems; accordingly, the metamaterial can achieve unique macroscopic EM properties that natural materials do not possess. These EM properties, such as equivalent permittivity and permeability, can be controlled by modifying the shape, size, and arrangement of meta-atoms in space. Such a new type of artificial EM medium can be applied to effectively control the propagation behavior of EM waves and bring about new physical phenomena. 

During 2011 and 2012, the concept of metasurfaces was first proposed, and the generalized law of metasurfaces was found in \cite{yu2011light}. Thereafter, EM materials changed from 3D objects towards planarization. The propagation characteristics (including the reflection, refraction, and diffraction) of EM waves can be effectively regulated by adjusting the amplitude and phase distribution of the metasurface. However, the metasurface, which is solidified by the traditional EM material, cannot regulate the EM wave in real time. In 2014, another group of researchers developed metamaterials in the direction of coding, digital, and programmable properties in \cite{cui2014coding}. The phase response of the meta-atoms over the metasurface can be encoded through PIN diodes and Field-Programmable Gate Array (FPGA) into zero or one in real time, thus transferring the design of metamaterial from the analog domain to the digital domain, building a bridge between the EM world and the information world, and further extending the concept of information metamaterial.  

RISs are concrete objects of information materials whose structure and geometric arrangement of meta-atoms can be reprogrammed according to the regulation of external signals. Specifically, RISs apply electronic phase transition elements (e.g. semiconductors or graphene) as switches or tunable reactance/ resistance elements between adjacent meta-atoms (or intra single meta-atom) to adjust the properties of individual meta-atom or the arrangement of a series of meta-atoms, thus realizing the function of reprogramming and reconfiguration. This type of digital and real-time metamaterials has constructed many system-level applications that are difficult to achieve by traditional metamaterials, such as the space-time coded digital communication system \cite{tang2020mimo,tang2020wireless,henthorn2019direct}and intelligent imaging and sensing systems \cite{li2019machine,li2019intelligent,li2020intelligent,hu2020reconfigurable}. 

\subsection{Fundamental characteristics of RIS}
Researchers in wireless communication community pay more attention to the characteristics of RISs at signal transmission level, such as how RISs can change and affect the signal during the propagation period.  Actually, by appropriately configuring the amplitude, phase, frequency, and polarization characteristics of the meta-atoms, RISs are capable of achieving various prospective uses.

\subsubsection{\textbf{Amplitude}}
In time-domain, numerous sub-wavelength sized meta-atoms with highly-controllable properties can be jointly tuned to determine how the incident signal will be like through the RIS. It has been already found in \cite{zhao2013tunable} that by adjusting the amplitude characteristic of the RIS, the RIS can absorb the undesirable signal (e.g. strong interference signals).  More specifically, within the designed frequency band, a unit cell in the tunable metamaterial absorber presents superior absorptivity covering a wide range of incident angles by regulating the bias voltage on the varactor diodes.

\subsubsection{\textbf{Phase}}
By adjusting the phase characteristic of the RIS, the spatial phase gradient is introduced so that the incident EM wave is no longer along the direction of the mirror image, according to which, both a single beam and multiple beams can be formed with the reflected direction being controlled freely. At present, the phase regulation characteristic is the most utilized for solving the problems in wireless communications, such as blind spots supplementary,  wireless communication links replenishment, system performance improvements, and so on. 

Firstly, RISs can provide supplementary links to compensate severe path loss and the channel sparsity, enriching the effective links between the base station (BS) and the user. Meanwhile, reflection coefficients, such as phase shift matrix can be jointly optimized with active beamforming at transmitters to improve the system performances including the spectral/energy efficiency, received signal-to-noise ratio (SNR) of the end-to-end link and so on. Secondly, by deploying RISs around the destination, the reflected signal can be focused on the particular spatial object by multi-beam focusing through reflected elements, gaining the equivalent performance compared to the conventional lens array. Moreover, an RIS can be regarded as a low-cost and lightweight alternative to large-scale MIMO systems with complex structures in millimeter-wave (mmWave) wireless communication since the whole system architecture possesses quite high hardware complexity and cost, as well as high power consumption whether applying analog beamforming or digital hybrid beamforming in large-scale MIMO systems, leading to more complicated algorithms and framework designs. 

\subsubsection{\textbf{Frequency}}
The frequency domain regulation of the RIS is to utilize the fast time-varying EM characteristics of RIS to control the frequency spectrum distribution. Specifically, EM waves can achieve non-linear spectrum shifts and allied functions of  radio frequency (RF) devices through the unique EM response of RIS, similar to the application of mixer and frequency multiplier. In this way, the spectrum characteristics of the signal can be expanded and adjusted, and the harmonics can be regarded as independent channels and control the amplitude and phase independently, so that each harmonic can transmit information independently, which taps the potential of RIS in the frequency domain  \cite{dai2020high,dai2020arbitrary}.

\begin{figure*}[t]   
	\centering  
	\includegraphics[height=7.5cm,width=18cm]{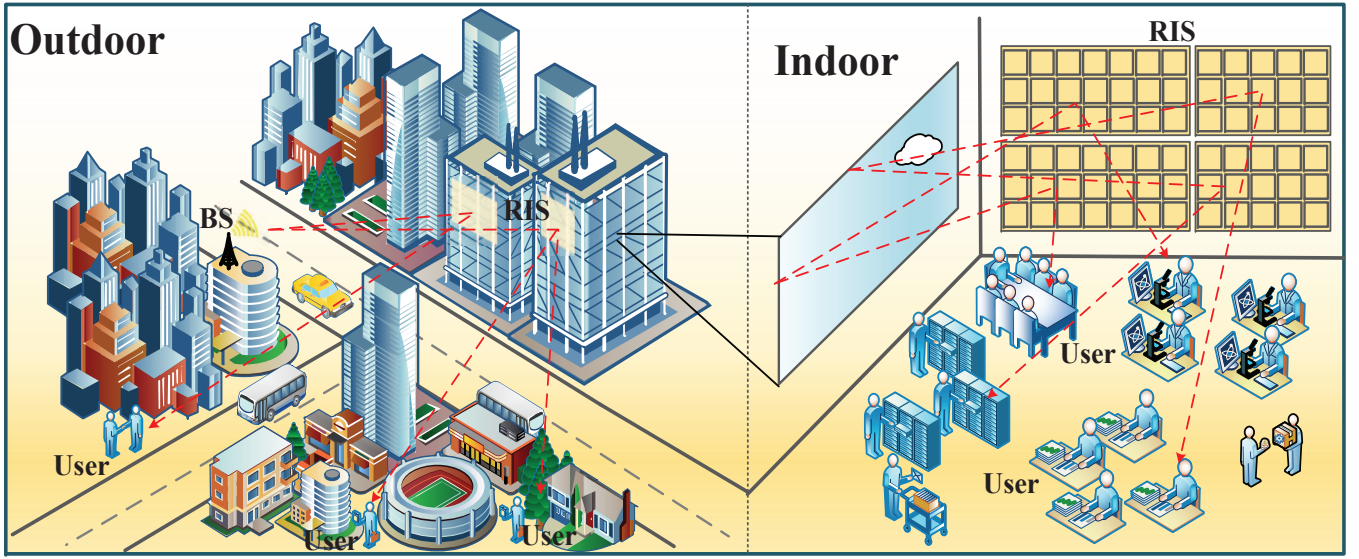}  
	\caption{RIS-assisted wireless communications for outdoor and indoor deployments. }  
	\label{img3}  
	\vspace{-0.2cm}
\end{figure*} 

\subsubsection{\textbf{Polarization}}
Besides, RISs can be deliberately designed to be polarization-sensitive. In the applications of RISs, the polarizations in different directions have varying amplitude or phase responses with high isolation. Polarization in each direction can be independently controlled through individual interface in real time, and the waves can carry different information, which gains the potential of multiplexing. To keep up with the surge studies of dual-polarized RISs \cite{liu2016anisotropic,zhang2020polarization,kelinear}, we have first explored the dual-polarized RIS-assisted wireless communication system and evaluated its performance, including ergodic spectral efficiency and optimal phase shift design considering practical hardware imperfect of the RIS. 

\subsection{Research status of RIS-assisted wireless communication}
RISs can be deployed in multiple scenarios in wireless communication networks in outdoor and indoor environments. In Fig.~\ref{img3}, RISs can provide extra links in the outdoor scenario when the direct paths between the BS and the users are blocked. In the indoor scenario, RISs can be regarded as a lightweight solution to rapidly increase the system capacity. Over the past several years, considerable tutorials and surveys of RIS have emerged. In \cite{9140329}, a comprehensive overview of smart propagation environments, RIS state-of-art research, RIS research projects and future research road ahead has been provided. In \cite{gong2019towards},  a literature review of the framework and application aspects of RIS has been presented. In \cite{ntontin2019reconfigurable}, the similarities, differences, advantages and disadvantages between the RIS and active relays have been exhaustively discussed. A comprehensive overview of RIS has been provided in \cite{wu2020intelligent}, wherein the fundamentals of RIS, recent research results, challenges, and potential for future research have been elaborated. In \cite{el2020reconfigurable}, a comprehensive tutorial on RISs for wireless communication is provided. In the aforementioned research, the channel modeling affected by the RIS implementation from two aspects of channel distribution and large-scale path loss, and challenges to be tackled in RIS optimization problems are concentrated and discussed in detail. 

The prototypes of RIS-based multi-stream transmitters have also received great interest \cite{tang2020mimo,tang2020wireless,henthorn2019direct}. A novel RIS-based wireless transmitter gains the comparable performance as the conventional channel coding methods with cost-effective and hardware-concise architecture without any filter, mixer, power amplifier, or RF chains. Moreover, path loss and channel modeling is essential for practical deployments of RISs. In \cite{tang2020}, free-space path loss models under far/near field beamforming and near field broadcasting scenarios have been proposed and validated by solid numerical simulation results. 

Performances enhancement is also one of the popular research topics that has been extensively studied in RIS-based wireless communication systems. Numerous papers emphasize on the RIS phase shift design, especially joint active beamforming design at the BS and passive beamforming design at RISs to achieve their goals for power consumption minimization, and spectral/energy efficiency improvement. For instance, reference \cite{wu2019intelligent} tackled the minimization problem of power consumption at the BS in a conventional convex optimization approach. Reference \cite{han2019large} obtained a closed-form solution of the optimal phase shift design for the RIS-assisted large-scale multiple-input single-output (MISO) system based on the derivation of the upper bound of the ergodic spectral efficiency and statistical channel state information (CSI).

In particular, channel estimation is an indicating  research topic because the CSI acquisition problem of RIS is challenging due to its nearly-passive properties. However, a few research studies have attempted to solve the problem. In Reference \cite{mishra2019channel}, the authors have proposed a channel estimation scheme on the basis of ON/OFF mechanism, which can improve the normalized mean-squared error (MSE) in single-user and multi-user scenarios. In  \cite{wan2020broadband,he2019cascaded}, low-overhead channel estimation algorithms have been proposed by taking advantage of the channel sparsity.

\subsection{What will it be like if RISs are embedded with AI?}
Conventional approaches for optimal RIS-assisted wireless communication network design have gradually presented various limitations, such as complicated establishment of the optimization problems, numerous slow optimal iterations, and high computational complexity. Accordingly, research topics towards AI-embedded RIS-assisted communications, have recently attracted considerable interest to break the above-mentioned original limitations. As illustrated in the previous sections, AI techniques are expected to be embedded into the RIS-assisted wireless communication networks for maximizing the availability of RIS to further realize the concept of smart radio propagation environments.

On the one hand, we can configure multiple sensors surrounding RISs to help capture the current range of environmental information, enabling RISs to seize large amounts of data that can be stored and locally processed, or retransmitted to the central controller for centralized processing in the BS. Accordingly, RISs configured with sensors can act as a part of the architecture of AI because the large amount of data obtained by RISs is necessary for training machine learning models and implementing data-driven machine learning  algorithms \cite{zappone2019wireless}. 

On the other hand, RIS phase shift matrices can also be optimized in addition to transceiver terminals compared with the traditional communication networks. Considering that multiple RISs will be in practical deployment and individual RIS is always configured by dozens or even hundreds of elements, RISs become particularly challenging in the phase shift design because the optimization of RIS-assisted wireless communication networks requires mathematical and considerable algorithm iterations, which is likely to gain time and computational complexity. In the scenario with high mobility, the complexity of the inter-connected smart devices in the IoT is increasingly difficult to deal with by utilizing stiff and static communication networks. Under such a condition, AI has become a key strategy for processing and integrating the data. Such a data-driven scheme can extract system features without requiring a concrete mathematical model\cite{ali20206g,letaief2019roadmap}. Thus, this mechanism can be applied to significantly simplify non-convex optimization tasks while making the training model robust against the various imperfections and, quickly adapt to the time-varying characteristics of the channel.

The AI schemes must be integrated into RIS-based networks to continue making further progress towards IWPE, making IWPE more feasible and robust to the non-linear factors in the system. In the following sections, we will illustrate how AI will be integrated into RISs and elaborate on various specific cases of study.

\section{Intelligent structures of metasurfaces}
Among various AI techniques, deep neural networks (DNN), also called multi-layer perceptron in deep learning (DL), utilize a hierarchical interconnection structure to imitate the connections between human brain neurons, which have been widely implemented for complex optimization problems. A DNN consists of an input layer, hidden layers, and an output layer. In this section,  two types of RIS based deep neural networks (Rb-DNNs) are introduced. The first one models programmable wireless environments (PWEs) as a DNN with RISs as neuron nodes. The other one models multi-layer digital-coding metasurface arrays as a complex valued DNN with meta-atoms as neuron nodes.

\subsection{Neural network-configured PWE (NN-CONFIG PWE)}
\begin{figure}[t]   
	\centering  
	\includegraphics[height=11.5cm,width=8.5cm]{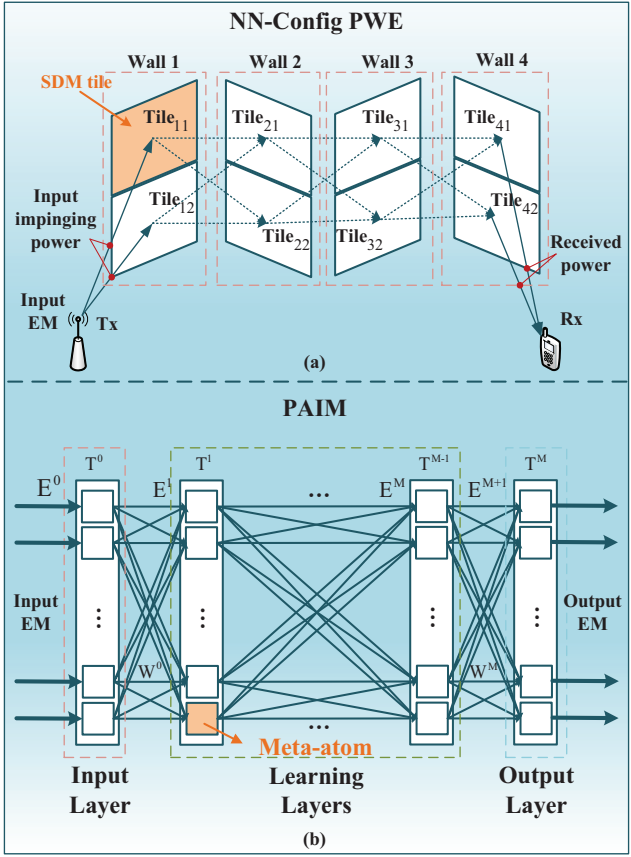}  
	\caption{ Paradigms of two intelligent structures of metasurfaces. (a) Conceptual framework of PWE as a neural network \cite{liaskos2019interpretable}. (b) 2D structure of the programmable AI machine (PAIM) model. ${{\rm{T}}^{\rm{i}}}$ represents the complex transmission coefficients, ${{\rm{E}}^{\rm{i}}}$ represents the EM field, and ${{\rm{W}}^{\rm{i}}}$ represents the space attenuation coefficients.}  
	\label{img4} 
	\vspace{-0.3cm}
\end{figure}

To possess deterministic and adaptive control over the EM wave propagation, the concept of PWEs has been proposed in \cite{liaskos2019interpretable}, which models them as an interpretable back-propagating neural network. In particular, this mechanism models software-defined metasurfaces (SDMs) as the neural nodes and their cross-interactions as links. A SDM tile contains multiple meta-atoms, and a layer has multiple SDM tiles, as illustrated in Fig.~\ref{img4}(a)). 

Specifically, a PWE consists of a transmitter (Tx), a receiver (Rx), and several walls deployed with SDMs. The incident EM waves from the Tx impinge upon the first wall (the first SDM layer), where the input layer units are configured by the propagation environments, such as Tx/Rx locations, densities, and dimensions of the RIS elements, frequency spectrum, and noise levels. Thus, an individual unit has its own impinging power, which can be split and reflected to the elements over the next layer; thus, cross-interactions are established as LoS links. The received power from the nodes in the last layer can be considered as the receiver output. The ideal output is the distribution over the Rx links, which can be obtained from the corresponding receiving gains of the Rx devices, as derived by the Rx antenna patterns and the MIMO configuration. The gap between the received and the ideal output is a metric for back-propagation period. After model training, the SDM tile-constructed neural network grasps the EM wave propagation characteristics and gains the capability of elevating communications between the BS and the user.

In another comprehensive work of \cite{liaskos2019interpretable} in  \cite{liaskos2020end}, the previous 2D precursor has been extended to operate in full 3D settings. 

\subsection{PAIM}
Previous research applies SDM tiles as neuron nodes to construct an interpretable neural network for adaptively configuring the PWE. Another work based on the fabricated all-optical diffraction deep neural networks ($\mathrm{D}^{2} \mathrm{NN}$) applies multi-layer digital-coding metasurface arrays as deep neural networks for implementing various tasks, such as image classification and EM wave sensing, even acting as a novel wireless communication coder - decoder or real-time multi-beam focusing instrument \cite{cui2020programmable}. The 2D structure of the PAIM model is shown in Fig.~\ref{img4}(b). It is a fully connected (FC) complex-valued neural network. The multiple squares represent meta-atoms in the RIS, whose complex transmission coefficients ${{\rm{T}}^{\rm{i}}}$ constitute the trainable part of the model. The EM field is represented by ${{\rm{E}}^{\rm{i}}}$, which can be attenuated according to the space attenuation coefficients ${{\rm{W}}^{\rm{i}}}$ and transmitted by the meta-atoms to all meta-atoms in the next layer.

The proposed PAIM can not only process typical DL tasks, such as image recognition and feature detection, but also act as a communication transceiver to manipulate the spatial EM waves and execute multi-channel coding and decoding or multi-beam focusing, which provides potential applications in the wireless communications, remote control, and other intelligent applications. 

\section{Intelligent systems of metamaterials}
As previously mentioned, the fast imaging, high-resolution sensing, and high accuracy localization will coexist with basic wireless communication functions in 6G, which can constantly share the abundant data and information in the time, frequency, and space domain. For instance, simultaneous sensing and mapping methods significantly enhance miscellaneous truly immersive XR services, and they can also be regarded as auxiliary approaches for autonomous systems, including vehicle navigation and drone cruise. Moreover, intelligent context-aware networks in 6G can also utilize localization and sensing information to optimize wireless resource allocation and execute appropriate operation with no or limited human intervention \cite{bourdoux20206g}. 

Particularly, RISs have the potential to obtain such imaging, sensing, and recognition capabilities via an appropriate design. In the following subsection, RIS-based intelligent systems are introduced in detail, ranging from the imaging, recognition, and sensing systems based on DL schemes to the adaptive intelligent metamaterial system based on AI.

\subsection{Imaging system}
Conventional microwave imaging systems need time-consuming data acquisition and complex reconstruction algorithms because they are based on a compressed sensing (CS) method and require iterative operations for data processing. Therefore, a tradeoff exists between imaging reconstruction speed and image quality. Given this background, the modern society is seeking for efficient and concise imaging systems, which are expected to quickly, intelligently, and efficiently reconstruct the image and extract important features with high fidelity and compression ratio.

To achieve fast super-resolution imaging, the feature information for the scenarios must be extracted in advance to apply some measurement modes. In \cite{li2019machine}, a super-resolution imaging system applies machine learning algorithms, called principal component analysis (PCA), to obtain features for scenarios and utilize information metamaterials to generate specific radiation patterns that match the scenarios. 
After a portrait is conducted, the PCA method is exploited to extract a series of feature maps of the portrait. Next, the information metamaterials will quickly form 400 radiation graphs based on 400 coding forms according to the extracted feature maps in each imaging period. The radiation graphs are connected with metamaterial coding forms through the discretized Gerchberg-Saxton (G-S) iterative algorithm. Then, a receiving antenna is deployed to receive the intensity of reflected waves after each radiation graph is irradiated to the human body. This process can obtain 400 sets of radiation graphs encoded by different metamaterial units and their corresponding mapping sets of reflected wave intensity. 

In the above work, the intensity of the reflected wave corresponding to the radiation graph obtained by each code is assumed to be proportional to the overlapping area of the object. The system equations can be established through the relationship between the portrait and the data pair, and portraits can be reconstructed according to these equations. In comparison, with those randomly generated radiation maps without using machine learning to acquire scenario information, the imaging performance of information metamaterial coding based on the PCA method for obtaining prior information is better than random radiation map imaging, clearly showing the superior performance of using machine learning and the importance of introducing machine learning algorithms. 

\begin{figure}[t]   
	\centering  
	\includegraphics[height=12.5cm,width=8.5cm]{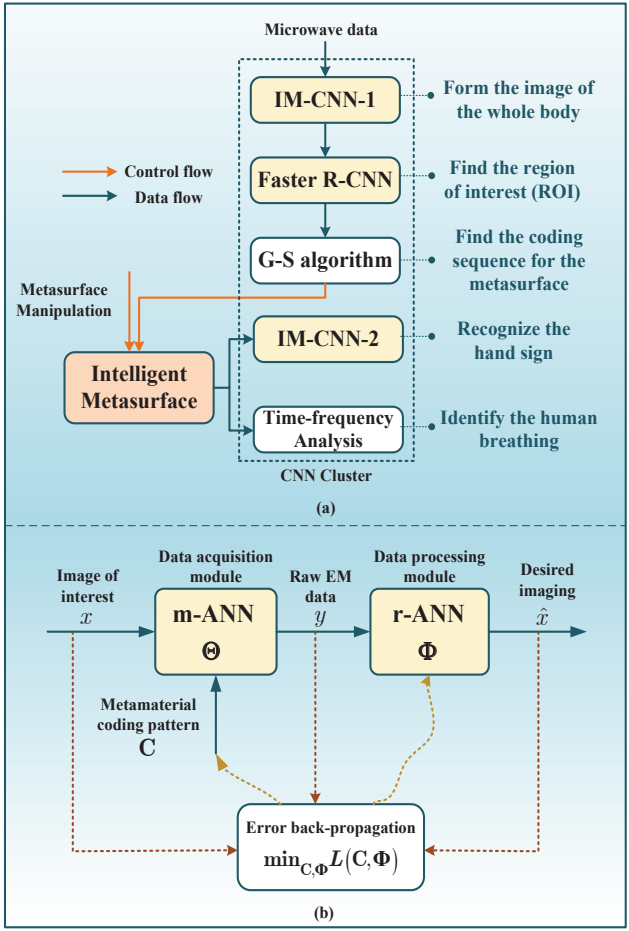}  
	\caption{ (a) Imaging and recognition system utilizing the DL-CNN cluster \cite{li2019intelligent}. (b) Intelligent sensing system based on the m-ANN data acquisition and r-ANN data processing module \cite{li2020intelligent}.}  
	\label{img5} 
	\vspace{-0.3cm}
\end{figure}

\subsection{Imaging and automatic recognition system}
The above imaging system in \cite{li2019machine} only introduces the PCA method in the acquisition of prior information of portraits while a traditional algorithm is still applied in the imaging period. In \cite{li2019intelligent}, the prior information of portrait is further extended to the imaging algorithm, thus forming a complete imaging system driven by the DL approach. As an extension, the system can execute extra DL tasks, such as gesture recognition and respiratory monitoring, as shown in Fig.~\ref{img5}(a). 

First, an end-to-end convolutional neural network (CNN) from the imaging input data to imaging target output is trained by IM-CNN-1. The training data are collected by shooting videos, and 80,000 training portraits are gathered. The PCA method is adopted to acquire feature information of portraits as \cite{li2019machine}, which is further organized as a 2D matrix as the input of the imaging system. With regard to the output, the portrait is extracted from the background of the image after the original image is collected with the optical camera and binarized as the target output of the imaging system. 

Next, the great advantage of DL has been demonstrated in the field of target recognition to extend the imaging system to gesture recognition and respiratory monitoring. Specifically, for the gesture recognition function, the hand position is located by Faster R-CNN according to the portrait formed by the system and the G-S algorithm is utilized to design the encoding of the information metamaterial to focus the EM wave energy on the hand, deploy the receiving antenna and collect the reflected wave intensity under different gestures as training data. Finally, a CNN, capable of recognizing 10 types of gestures and with a recognition accuracy of 95$\%$ is trained by IM-CNN-2. In the respiratory monitoring task, the same method is applied to locate toward the chest cavity, collect echo data for a period of time, and perform time-frequency analysis to obtain the target's respiratory status during that period.

\subsection{Intelligent sensing system}
Although the imaging and recognition systems in \cite{li2019intelligent} introduced the DL methods, including CNN and R-CNN, into the data acquisition and imaging period, these two periods must be independently trained and require a large number of training samples. In \cite{li2020intelligent}, the neural network that combines the data acquisition process with imaging process is proposed with information metamaterial coding enrolled. 

Specifically, the intelligent sensing system consists of two data-driven modules, namely, the m-ANN data acquisition module and the r-ANN data processing module, as illustrated in Fig.~\ref{img5}(b). The input of the m-ANN network (equivalent to the encoder) is the image of interest \textit{x} and the coding pattern of the metamaterial \textbf{C}. The output is the received raw microwave data \textit{y} collected by the receiving antenna, which is also the input of the r-ANN network (equivalent to the decoder). The target output of r-ANN is the desired imaging of interest $\hat{x}$. First, the weight of m-ANN $\mathbf{\Theta}$ is fixed by supervised learning. Next, the metasurface coding pattern \textbf{C} and the weight of r-ANN $\mathbf{\Phi}$ are jointly learned by error back-propagation. The proposed network connects the entire imaging system into a whole, and the author can find optimal metamaterial radiation patterns (codings) with less number of radiation pattern image requirement through the joint training while ensuring the imaging quality. 

\subsection{Adaptive intelligent metamaterial system}
Reconfigurable metasurfaces have been developed to dynamically and arbitrarily manipulate the EM waves to execute variable tasks. However, manual controls are required to switch among different functionalities. Therefore, the adaptive intelligent metamaterial system is proposed in \cite{ma2019smart}, which is a closed-loop decision-making system integrated with sensors, feedback links, and algorithms to ensure that the programmable metasurface can self-adaptively switch between different functions without manual adjustment.

When sensing the changes of the outside environment, the sensed information is fed back to the FPGA and microprocessor for analysis. Then, the closed-loop decision system executes the corresponding metamaterial code distribution according to the internal preset algorithm to achieve beam control. The metasurface can adjust the EM radiation beam according to its motion posture; accordingly, the communication beam always points to the satellite. In addition to the beam staring function, the metasurface can also implement other functions such as multi-beam dynamic tracking, dynamic RCS scattering control, vortex wave generation, and multi-beam scanning.

\begin{figure}[t]   
	\centering  
	\includegraphics[height=6cm,width=8.5cm]{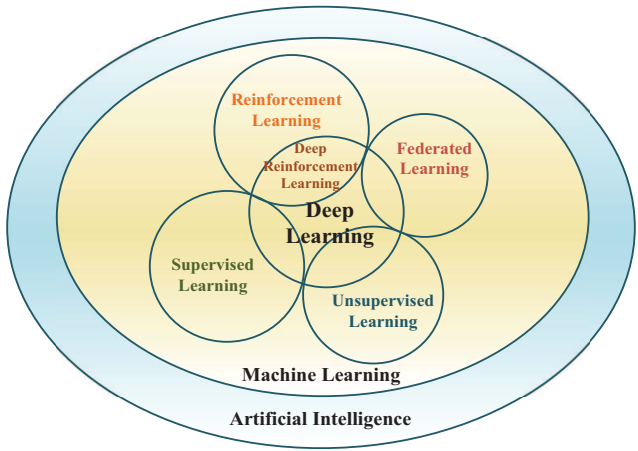}  
	\caption{ Relationship among AI, machine learning, supervised learning, unsupervised learning, DL, reinforcement learning (RL), and federate learning (FL) \cite{li2017deep}.}  
	\label{img6} 
	\vspace{-0.3cm}
\end{figure}

\section{AI-embedded RIS-aided wireless communication system}

The recent years have witnessed the rapid development of AI in financial analysis, e-health care, and industry 4.0. The future communication networks will also have to rely on AI. Some basic concepts still need to be clarified. The relationship among AI, machine learning, supervised learning, unsupervised learning, DL, reinforcement learning (RL), and federate learning (FL) is illustrated in Fig.~\ref{img6} and the basic concepts of machine learning schemes are presented in Table~\ref{I}. 

The DL method utilizes multi-layer non-linear mapping networks via hierarchical connections of like-brain neurons and has potential to efficiently solve optimization problems. Recently, DNNs have been shown to provide superior performance in various tasks such as image recognition and natural language processing. Moreover, CNNs further play a critical role in DL because the convolution kernel parameter sharing in the hidden layer and the sparsity of the connections between layers enable the CNN to perform and learn grid-like topology features with a small amount of calculation. 

In addition to the DL, the RL \cite{luong2019applications} utilizes an agent to interact with the environment and learns how to take actions in the next state. At each step of the learning process, the agent observes the current state of the environment and generates an action. After the agent receives a reward, it moves to the next stage. In deep RL (DRL), the critic and actor networks take DNNs as the main structure. Given that the agent aims to maximize the long-term cumulative rewards, many wireless communication problems, such as resource allocation, can be formulated as a RL problem. 

The FL \cite{niknam2020federated} is a distributed machine learning algorithm that enables mobile devices to collaboratively learn a shared machine learning model without frequent data exchange between mobile devices and servers. In the FL framework, each mobile device and its datacenter owns independent machine learning models, called the local FL model, and the machine learning model of the datacenter is called a global FL model. 

\begin{figure*}[t]   
 	\centering  
 	\includegraphics[height=10cm,width=18cm]{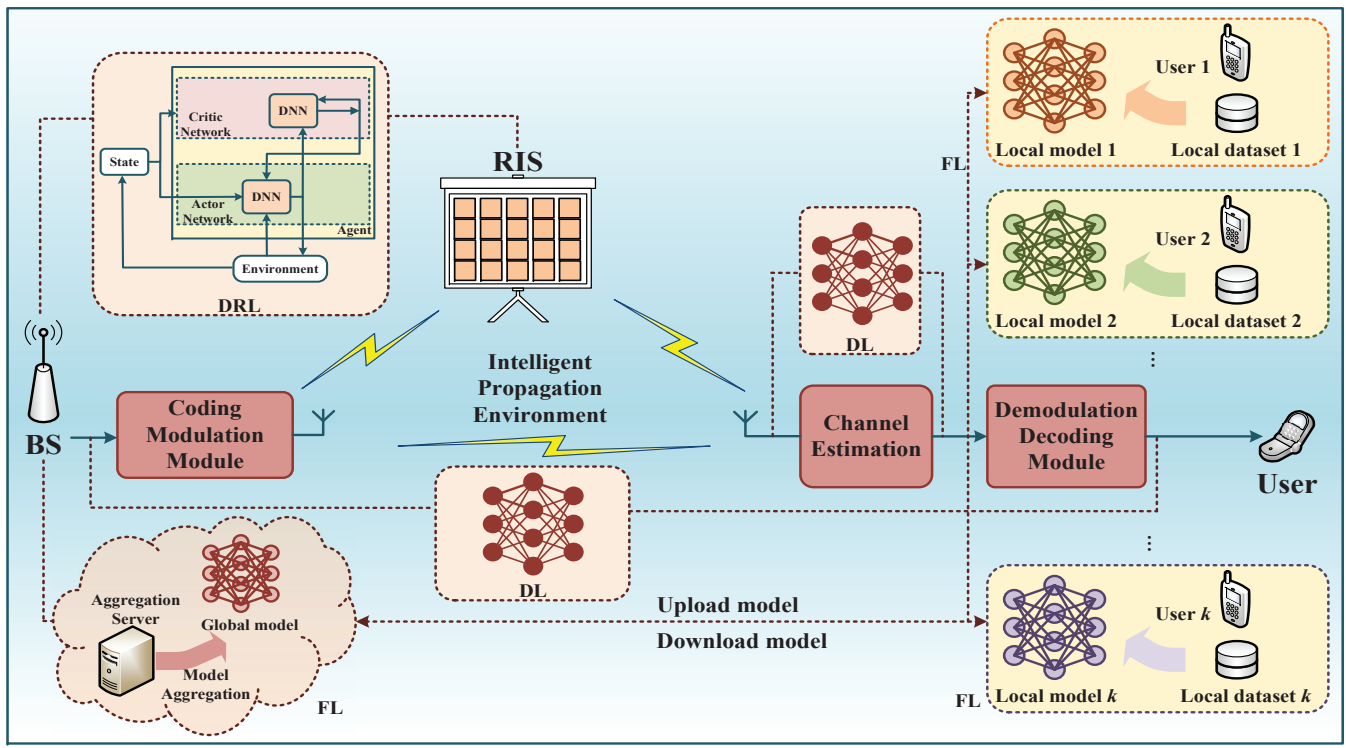}  
 	\caption{ Architecture of the AI-embedded RIS-assisted wireless communication system. }  
 	\label{img7}  
 	\vspace{-0.2cm}
 \end{figure*}

Specifically, supervised learning and unsupervised learning are two learning schemes that can be distinguished by the existense or absense of the output labels. In the following section, AI-embedded RIS-assisted wireless communication systems are elaborately overviewed  (Fig.~\ref{img7} ), including the signal detection, channel estimation, and beamforming design at the physical layer, resource allocation, and over-the-air computation (AirComp) at the upper layers.
\vspace{-1cm}

\begin{center}
	\begin{table}[t] 
		\caption{Basic concepts of machine learning schemes}
		\label{I}
		\renewcommand\arraystretch{1.5}   %调整表格行高
		\begin{tabu} to 0.5\textwidth{p{2.5cm}|p{6cm}}  
			\hline  
			Schemes & Characteristics   \\ 
			\hline  
			DL & A model utilizing multi-layer non-linear mapping networks via hierarchical connections of like-brain neurons. DNNs and CNNs are most widely used especially in DRLs for state
			representation, function approximation for value function, policy, transition model, or reward. \\  
			\hline  
			RL & A model contains an agent, in which the agent interacts with the environment, learning an optimal policy, by trial and error, for maximizing the long-term cumulative rewards. \\ 
			\hline 
			FL & A distributed machine learning algorithm that enables geo-distributed devices to collaboratively learn a global model while keeping the data processed locally.\\ 
			\hline  
			Supervised Learning& A scheme that training the model by illustrating input samples and their known associated output labels.    \\  
			\hline  
			Unsupervised Learning& A scheme that  the model learns to classify input samples optionally without given output labels.    \\
			\hline  
		\end{tabu}  
	\end{table}  
\end{center}

\begin{center}
	\begin{table*}[t] 
		\centering
		\caption{Channel Estimation}
		\label{II}
		\renewcommand\arraystretch{1.7}   %调整表格行高
		\begin{tabu} to 0.5\textwidth{p{1cm}|p{2cm}|p{14cm}}  
			\hline  
			Ref.  & Model & Highlights and drawbacks   \\ 
			\hline  
			\cite{elbir2020deep}& double CNNs &  The RIS is completely passive, and double CNNs estimate the direct channel and cascaded channel, respectively. However, there is need for extra control links and signals for the manipulation of RIS elements (turn on/off).\\  
			\hline  
			\cite{liu2020deep}& CV-DnCNN & Combination of model-driven and data-driven: Conventional CS channel estimation and reconstruction for rough estimation and CV-DnCNN model for accurate channel estimation. However, a hybrid passive/active framework requires extra power consumption and control signals for the manipulation of RIS elements. \\
			\hline  
			\cite{jiang2020channel}  & DNN & Combination of model-driven and data-driven: Direct calculation (DC) for rough estimation and DNN model for accurate channel estimation. The hybrid passive/active framework gains high complexity as well. \\  
			\hline  
			\cite{liu2020deepdeep}  & CDRN  & To improve the estimation accuracy, the proposed framework take fully advantages of CNN in feature extraction and DReL in denoising. \\ 
			\hline 
			\cite{elbir2020federated} & FL-CNN & The FL scheme avoids huge transmission overhead, and a single CNN is trained for two different datasets for both direct and cascaded channels. \\ 
			\hline  
			\cite{Xu2020}  & ODE-based CNN  & The proposed structure not only take the advantage of the CNN in using the correlation between elements for information completion, but also introduce the ODE to describe the latent relationship between different layers to speed up the convergence and the learning performance. \\ 
			\hline 
		\end{tabu}  
	\end{table*}  
\end{center}

\subsection{Survey of RIS with AI}

In the survey in \cite{zappone2019wireless}, AI is considered to play a critical role in RIS-assisted wireless communication systems as a supplement to the traditional mathematical model-based wireless network design. In \cite{zappone2019wireless}, interactions is said to occur between the AI-based wireless communication networks and the RIS-based smart wireless radio environments. The RISs can acquire environmental information by configuring various sensors, and the data-driven characteristic of AI can reduce the high computational complexity of the optimization design of RIS-based network with hybrid active points and numerous passive RIS elements. 

According to another comprehensive survey about RIS in \cite{9140329}, machine learning must be applied into the RIS-assisted network to realize a truly smart radio environments because machine learning methods can perform intelligent tasks. This notion means that these methods can intelligently interact with the wireless radio environment by planning, perceiving, reasoning, learning, and problem solving.

Another work presents a concise introduction of \textit{Wireless 2.0} in \cite{gacanin2020wireless}, which means the intelligent radio environment leveraging RISs and AI. This research focuses on the AI-based computational methods and elaborates the application of AI in \textit{Wireless 2.0} from the perspective of classification. In \cite{elbir2020survey}, a short survey of DL technologies utilized in RIS-assisted wireless systems is presented but, without considering those intelligent structures and systems of RIS.

\subsection{Physical layer}

\subsubsection{Signal detection}
\ 

In \cite{khan2019deep}, an RIS-assisted MISO communication system is considered, and a novel detector, called “DeepRIS” is developed for estimating and detecting symbols signals transmitted through RIS on the basis of a DNN framework. Specifically, the RIS is fully passive to make it as cost-efficient as possible instead of deploying active elements on RIS to assist detection. 

The proposed DeepRIS, acting as a data-driven black box, consists of three FC layers. The model is trained offline by utilizing different simulated channels and RIS phase shift matrices and, a large amount of training patches, including the transmitted vectors acting as the output and the received vectors as the input. After training, the model can directly estimate the transmitted symbols beyond the complicated step of channel estimation in RIS, thus reducing the pilot signaling overhead. 
Moreover, the bit-error-rate of the proposed method outperformed that of the traditional detectors, such as least squares and minimum mean-squared error (MMSE) estimators, which can achieve a near-optimal bit error rate by maximum likelihood (ML) estimator.
 
\vspace{0.3cm}
 \subsubsection{Channel estimation} 
 \ 
 
The channel estimation of RIS-based wireless communication systems is faced with  much more severe challenges than that of the conventional communication scenarios. On the one hand, RISs do not require a large amount of RF chains due to the passive reflection characteristics. In comparison with MIMO and relay with strong signal processing capability, RISs are only equipped with simple on-board signal processing units. On the other hand, each RIS consists of a huge number of reflecting elements, leading to the challenges for CSI acquisition (if the CSI acquisition is possible). 

To carry out an effective channel estimation, new algorithms and protocols must be designed to avoid the complex on-board signal processing operations. Moreover, the near-field/ far-field channel modeling of RISs have different propagation characteristics, and the introduction of RISs may affect the EM field. These challenges hinder the characterization and simplification of cascaded BS-RIS-UE channels. 

Reference \cite{elbir2020deep} pioneered a DL framework for channel estimation in RIS-assisted mmWave massive MIMO systems. A twin CNNs called \textit{ChannelNet}, has been proposed to estimate the channels, where the direct channel (BS-UE) and the cascaded channel (BS-RIS-UE) are estimated by configuring two types of phases for the pilot training model. In phase I, all RIS elements are turned off for the direct channel estimation. And in phase II, the RIS elements are turned on one by one or they are turned on simultaneously for the cascaded channel estimation. 

However, the model training of the DL-based method is conducted in a centralized manner in \cite{elbir2020deep}, which poses much more training overhead. Moreover, in cascaded channel estimation, extra control links are necessary for controlling the on/off state of reflecting elements. To solve this problem, deep denoising neural networks can be effective. In \cite{liu2020deep}, a hybrid passive/active RIS architecture is first proposed, in which a small part of RIS elements are activated and a few receive chains are deployed to estimate the partial channels. A conventional CS algorithm, i.e., orthogonal match pursuit (OMP) is applied to reconstruct the complete channel matrix whereby the angle domain is sparse. After preliminary estimation, a complex-valued denoising convolution neural network (CV-DnCNN) is used to further enhance the estimation accuracy. Similar to \cite{liu2020deep}, the overall channel estimation problem is also divided into two tasks in \cite{jiang2020channel}. The first task is to activate a small number of RIS elements for the angle parameter estimation, and the second task is to utilize a DL framework for further estimation accuracy improvement. Instead of complicated CS, the direct calculation (DC) approach is applied to directly calculate the angle parameters through the channel estimation of the adjacent active elements. 

According to the simulation results, these hybrid architectures gain better performance by leveraging conventional channel estimation approaches and DL methods; however, they require extra power consumption and control signals for the manipulation of RIS elements.

Taking advantage of the feature extraction of CNN and the denoising capability of deep residual learning (DReL), reference \cite{liu2020deepdeep} proposed a CNN-based deep residual network (CDRN) to tackle the channel estimation, which is regarded as a denoising problem. This approach can intelligently exploit the channel spatial features and implicitly learn the residual noise for further improving the channel estimation accuracy.

The model training in \cite{elbir2020deep,liu2020deep,jiang2020channel,liu2020deepdeep} is  centralized in the BS, introducing huge transmission overhead from the users to the BS. FL-based channel estimation can solve the problem. Reference \cite{elbir2020federated} proposed a FL framework for channel estimation, wherein the learning model is trained at the local user with its own local dataset. In comparison with the centralized approaches, only model updates are sent rather than all datasets, thereby reducing the transmission overhead. Furthermore, a single CNN is trained for two different datasets for both channels instead of using double CNNs to estimate the direct channel and the cascaded channel in the previous work \cite{elbir2020deep}. 

In \cite{Xu2020}, an ordinary differential equation (ODE)-based CNN structure is proposed to extrapolate the full channel information from the partial channel. Specifically, the sub-sampled RIS channel is formulated by turning on a fraction of all the RIS elements. The ODE-based CNN, where the cross-layer connections are added to describe the latent relationship between different layers, is adopted to learn the mapping function from the channel of chosen elements to that of the all elements. The proposed CNN can achieve better performance than the cascaded CNN because extra coefficients and linear calculations are introduced.

These studies relevant to the channel estimation are concluded in Table~\ref{II}, including their machine learning models and highlights/drawbacks.

\vspace{0.3cm}
\subsubsection{Beamforming design for performance enhancement}
\ 

The RIS phase shift configuration is critical to the enhancement of the system performances. Nevertheless, the optimal beamforming design is a challenging issue due to the non-convex constraints on the RIS reflecting elements and the various complicated non-convex optimization objective functions. Conventional phase shift design approaches are mostly to find a sub-optimal solution (no closed-form solution is available) on the basis of the semi-definite relaxation (SDR) technique, for instance, in \cite{wu2019beamforming}, to maximize the received signal-to-noise ratio (SNR). Given that the SDR method is of high computational complexity, a relatively low complexity fixed-point iteration algorithm has been proposed in \cite{yu2019miso} for optimization problems. In \cite{wu2019intelligent}, the greedy manner is a promising solution for combating the high performance loss when the user is located far away from the BS, in which the phase shift of individual unit is iteratively optimized. The resulting sub-optimal iterative algorithms incur high complexity; hence, they are not suitable for real-time implementation. On this basis, the DL, RL, and FL methods have been introduced for the phase shift design in the RIS-assisted wireless communication systems.

Reference \cite{taha2019enabling} considered two different types of methods, namely, CS and DL, as efficient solutions for channel estimation and optimal RIS reflection matrix design. In the second task, a DL framework is utilized to help the RIS in optimally interacting with the incident signal and design its phase shift matrix to ensure that it can represent the current state of the environment and the transmitter/receiver locations, which maximizes the achievable rate of the system.

Reference \cite{huang2019indoor} addressed the RIS beamforming design problem in an indoor communications scenario. The proposed network employs a DL method to increase the received signal strength for the indoor user by training a DNN to learn the mapping between the user coordinates and the RIS phase shift configuration. After training from data (a preconstructed fingerprint database), the trained model can map the target user coordinate to the optimal phase shift matrix as the output, further focusing the beam to the target user location. However, this work still applies a hybrid RIS architecture wherein the RIS is also equipped with several activated elements for channel estimation, which blurs the passive nature of the RIS. A further adaptability analysis of the proposed network is desired because the current analysis assumes that only the perfect CSI condition was considered.

Supervised learning is extremely dependent on a large amount of labeled data in advance for model training. An efficient way to leapfrog the labeling acquirement process is to utilize label-free unsupervised learning techniques. Reference \cite{gao2020unsupervised} also proposed a DL approach for passive beamforming design. In contrast with the previous work \cite{huang2019indoor}, a standard DNN that comprised five FC layers is trained offline in terms of an unsupervised learning mechanism to avoid tremendous training labels overhead compared with the supervised learning-based DL approach. Fed with the product of the BS to RIS, RIS to users, and BS to user channels, the proposed method can predict the phase shift configuration as the output. Simulation results illustrate that this method can achieve comparable performance as conventional SDR-based approaches with much lower computation complexity.  

Previous work in \cite{huang2019indoor} has introduced current DNN-estimated channel information to design RIS reflection matrices. Acorrelation exists between the current and the previously estimated channels, which is ignored by many researchers. In contrast with the existing works, reference \cite{aygul2020deep} exploit the channel correlation to more reliably configure the RIS reflection interaction. The simulation results demonstrate the performance improvements achieved by introducing the strong correlation between the previously sampled channels and the ones being estimated.

\begin{figure}[t]   
	\centering  
	\includegraphics[height=11.5cm,width=8.5cm]{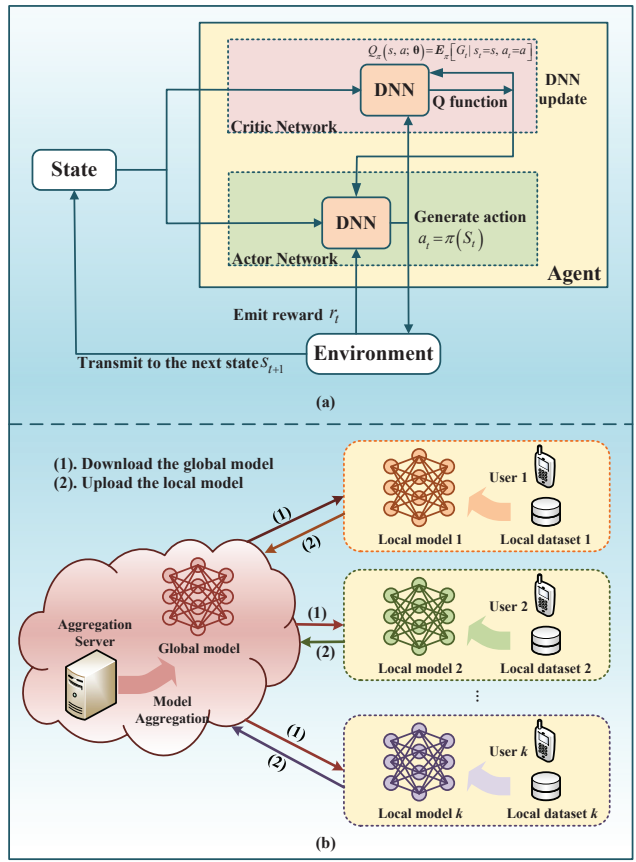}  
	\caption{ (a) Illustration of the DRL architecture, where double DNNs are utilized to approximate the optimal state-action value and Q function. (b) Illustration of the FL framework. Individual user utilizes its own local dataset to generate its local model and updates the model to the aggregation server. The server executes the model aggregation to generate the global model, which is sent back to the users.}  
	\label{img8}  
	\vspace{-0.2cm}
\end{figure}

Various DNN architectures are leveraged for designing RIS beamformers \cite{taha2019enabling, huang2019indoor,gao2020unsupervised,aygul2020deep}. To avoid huge training labels, DRL-based methods, which achieve the property of online learning and sample generation, are widely utilized. In \cite{feng2020deep}, an efficient DRL method is proposed to solve the non-convex optimization problem of the phase shift deign for the RIS-aided downlink MISO wireless communication system to maximize the received SNR and, in \cite{lin2020optimization}, to minimize the BS transmit power by jointly optimizing the active beamforming at the BS and passive beamforming at the RIS, as shown in Fig.~\ref{img8}(a). The deep deterministic policy gradient (DDPG) algorithm is introduced into the DRL framework. 

Reference \cite{huang2020reconfigurable} also investigated the joint beamforming design of transmit beamforming matrix at the BS and the phase shift matrix at the RIS by leveraging the recent advances in DRL with the model-driven DDPG approach. However, unlike the aforementioned work in \cite{lin2020optimization} that applies alternating optimization to alternatively obtain the optimal transmit beamforming and phase shift matrix, the proposed method can simultaneously achieve the optimal transmit beamforming and phase shift matrix by maximizing the sum rate, which is utilized as the instant rewards to train the DRL-based algorithm. The same group of authors also investigated the joint design of digital beamforming matrix at the BS and analog beamforming matrices at the RISs by leveraging the DRL framework to combat the propagation loss in \cite{huang2020hybrid}, which further shows that DRL-based architectures are those effective methods for tackling the non-convex optimization problems, such as NP-hard beamforming problems.

To eliminate the challenging labeling process of the supervised learning-based DRL techniques, reference \cite{taha2020deep} employed an unsupervised learning-based DRL approach to design the RIS reflection matrices. In the method, a few elements are activated to acquire channel information, and the state is updated according to the normalized concatenated sampled channel of each transmitter-receiver pair. Moreover, the achievable rate at the receiver is utilized as the reward to train the network. Specially, the proposed framework is said to be \textit{directed towards standalone RIS operation}. This notion means that the RIS architecture adapts and interacts with the surrounding environment for phase shift configuration rather than by controlling signal from the BS through the wired connection between the BS and the RIS.

Referencce \cite{gong2020optimization}, designed a novel optimization-driven DRL framework for the joint beamforming optimization problem, which takes the advantages of the efficiency in model-based optimization and the robustness in data-driven machine learning approaches. In \cite{lee2020deep},  a novel Deep Q-network (DQN) approach based on DRL is proposed, in which the BS receives the state information, consisting of the users CSI feedback and the available energy reported by the RIS, to maximize the average energy efficiency by enabling the BS to determine the transmit power and RIS configuration with uncertainty on the wireless channel and harvested energy of the RIS system.

The decaying DQN (D-DQN)- based algorithm proposed in \cite{liu2020machine} can tackle UAV trajectory and RIS phase shift design problem. In this algorithm, the central controller is selected as an agent for periodically observing the state of the UAV-enabled wireless network and executing actions for adapting to the dynamic environment. In contrast with the conventional DQN algorithm, the decaying learning rate is leveraged in the proposed D-DQN-based algorithm for a tradeoff between accelerating training speed and converging to the local optimum.  

The DRL and FL approaches can also be adopted to enhance the physical layer security and privacy. In \cite{9206080}, a joint active beamforming at the BS and passive beamforming at the RIS optimization problem considering the different quality of service (QoS) requirements  and time-varying channel conditions is formulated to improve the secrecy rate of multiple legitimate users. DRL solves the problem by formulating the reward function as the difference between the secrecy rate and a penalty term, accounting for the QoS at the receivers.

To consider user privacy, reference \cite{ma2020distributed} exploited the phase shift design of RIS. An optimal phase shift design based on FL is proposed with the sparse CSI to simultaneously enhance the data rate and protect the data privacy.
Several local models are trained according to the CSI of users and further uploaded to a central server for aggregation to generate a global model. The user can download the global model as the initial configuration for the next training period. 

Age-of-Information (AoI) is defined as the elapsed time since the generated/sampled of the most recently received status-update. To fully characterize the freshness of status-update, the concept of AoI has been introduced as a main performance metric for those applications that require reliability and timeliness in delivering status-update information, such as smart environmental monitoring, industrial control systems, and intelligent transportation systems. Reference \cite{samir2020optimizing} investigated a wireless network in which IoT devices (IoTDs)  with limited transmission capabilities need to sample the stochastic process and deliver the sampled data to a BS for processing, and an aerial RIS (deployed over a UAV) provides supplementary links for the sampled data delivering. To address the challenging problem, a DRL framework based on proximal policy optimization is proposed to effectively learn the IoTD activation patterns, control the UAV altitude, and find the optimal RIS phase shift design to minimize the expected sum AoI.

In summary, the studies relevant to the RIS beamforming design are listed in Table~\ref{III}, wherein their machine learning models and optimal objectives are included.

\begin{center}
	\begin{table}[t] 
		\caption{Beamforming design}
		\label{III}
		\renewcommand\arraystretch{1.5}   %调整表格行高
		\begin{tabu} to 0.5\textwidth{p{0.7cm}|p{1.6cm}|p{5cm}}  
			\hline  
			Ref.  & Model & Objective   \\ 
			\hline  
			\cite{taha2019enabling}& DNN &  Maximize the achievable rate of the system.\\  
			\hline  
			\cite{huang2019indoor}& DNN& Increase the received signal strength (RSS).  \\
			\hline  
			\cite{gao2020unsupervised} & UL-DNN & Enhance the effective gain of the reflecting path. \\  
			\hline  
			\cite{aygul2020deep}& DNN  & Maximize the achievable rate of the system.\\ 
			\hline 
			\cite{feng2020deep} & DRL-DNN &  Maximize the received SNR.\\ 
			\hline  
			\cite{lin2020optimization }& DRL-DNN & Minimize the BS’s transmit power. \\ 
			\hline  
			\cite{huang2020reconfigurable }& DRL-DNN & Maximize the sum rate of the system. \\ 
			\hline  
			\cite{huang2020hybrid} & DRL & Combat the propagation loss in terahertz band.\\ 
			\hline  
			\cite{taha2020deep} & UL-DRL &Maximize the achievable rate of the system. \\ 
			\hline  
			\cite{lee2020deep} & DQN & Maximize the average energy efficiency. \\ 
			\hline  
			\cite{liu2020machine} & DQN&  Tackle the UAV trajectory. \\
			\hline  
			\cite{9206080} & DRL & Improve the secrecy rate of legitimate users. \\ 
			\hline  
			\cite{ma2020distributed} & FL & Enhance the data rate and protect the data privacy simultaneously. \\ 
			\hline  
			\cite{samir2020optimizing}& DRL & Minimize the expected sum AoI. \\ 
			\hline 
		\end{tabu}  
	\end{table} 
\vspace{-0.3cm} 
\end{center}

\subsection{Upper layers}
Communication efficiency is essential in exploiting massive amounts of data generated at the mobile user equipment. FL has been widely utilized as a potential substitute for centralized ML schemes. It avoids heavy data transmission overhead by collaboratively illustrating a shared global model at the server, while the local data is processed at distributed services only. This method also reduces the communication latency and enhances the user privacy and data security, compared with the conventional centralized machine learning schemes. The FL framework is shown in Fig.~\ref{img8}(b)). When the local datasets become large and local models are complex, training machine learning models are prefer to distribute the model parameter optimization over multiple local devices and generate the global model at an aggregation server by model aggregation.

AirComp provides a novel simultaneous access technique to support fast model aggregation for FL via exploiting the signal superposition property of multi-access channels, which can be regarded as one of the uplink non-orthogonal multiple access schemes. However, AirComp suffers the severe propagation errors since the local parameters are transmitted over the shared wireless channels, thus deteriorating the global model aggregation performance. By contrast, RIS-aided FL can be as an innovative and promising candidate to boost the model aggregation process, effectively minimize the propagation error, and enhance the convergence rate of FL.  The available links for cell-edge users who are blocked by obstacles are also provided to expand the coverage and connectivity of FL, thus boosting the utilization of wireless network resources.

With the advancement of AI, Intelligent IoT (I-IoT) will be innovative, moving from connected things towards connected intelligence. Under such a condition, the FL method is an effective solution for those privacy sensitive and low-latency intelligent IoT services and applications, including autonomous cars and massive robotics. Reference \cite{yang2020federated} proposed an AirComp based communication-efficient FL framework for such intelligent IoT networks to  provide low-latency decisions with strong privacy and security guarantees for applications in the IoT network. In the proposed framework, the RIS is leveraged to reduce the model aggregation error for AirComp-based FL.

Particularly, the MSE is widely adopted as the performance metric to characterize the model aggregation error, which heavily depends on the channel conditions between the local devices and the aggregation server. Minimizing the model aggregation error of MSE quantization for AirComp-based FL is the key to improving the learning performance. Given that the RIS can tailor the IWPE to obtain the desired channel responses, the MSE of the global model aggregation is accordingly reduced, and the model prediction accuracy is also improved. Moreover, the smaller MSE makes it possible to select more local devices at each round of communication, thereby accelerating the convergence of FL.

Although the work in \cite{yang2020federated} has demonstrated the effectiveness of RIS for AirComp model aggregation improvement, it merely concentrates on the communication aspect. In \cite{liu2020reconfigurable}, an RIS-enabled FL system is considered, in which the RIS phase configurations and device selection are jointly optimized in a uniform framework to act on the FL performance. Specifically, an effective algorithm based on Gibbs sampling and successive convex approximation principle is proposed, which aims to jointly optimize the device selection, the
receiver beamforming, and the RIS phase shifts. Numerical experiments also verify that enhanced FL performance can be achieved by unifying the communication system design and user equipment selection under a unified framework.

Reference \cite{ni2020federated} also investigated the model aggregation problems in the multiple RISs-aided FL system. A novel framework of resource allocation and device selection is proposed for optimal global parameter aggregation with the help of multiple RISs. This work not only reduces the model aggregation error but also accelerates the convergence rate of the global aggregation in FL. This task is carried out by jointly optimizing the transmit power, receive scalar, RIS phase shifts, and learning participants selection subject to the constraints, including the transmit power for devices, the phase shift of RIS reflected element, and the aggregation requirement. The simulation results show that the proposed communication-efficient resource allocation algorithms with the aid of multiple RISs outperform the benchmarks (i.e., single RIS auxiliary, random phase shift of the RIS), specifically illustrated as better convergence rate and lower learning error.

\section{Challenges and potentials}
To further make IWPE a reality, some challenges must be addressed and potentials must be explored, including the novel wireless transceiver based on the RIS neural network, RIS-based localization and sensing system design, dynamic and flexible control of RIS, and the data collection and model training algorithm design in machine learning schemes.
\subsection{Towards the intelligent structure of RIS}

\begin{figure*}[t]   
	\centering  
	\includegraphics[height=7.5cm,width=18.5cm]{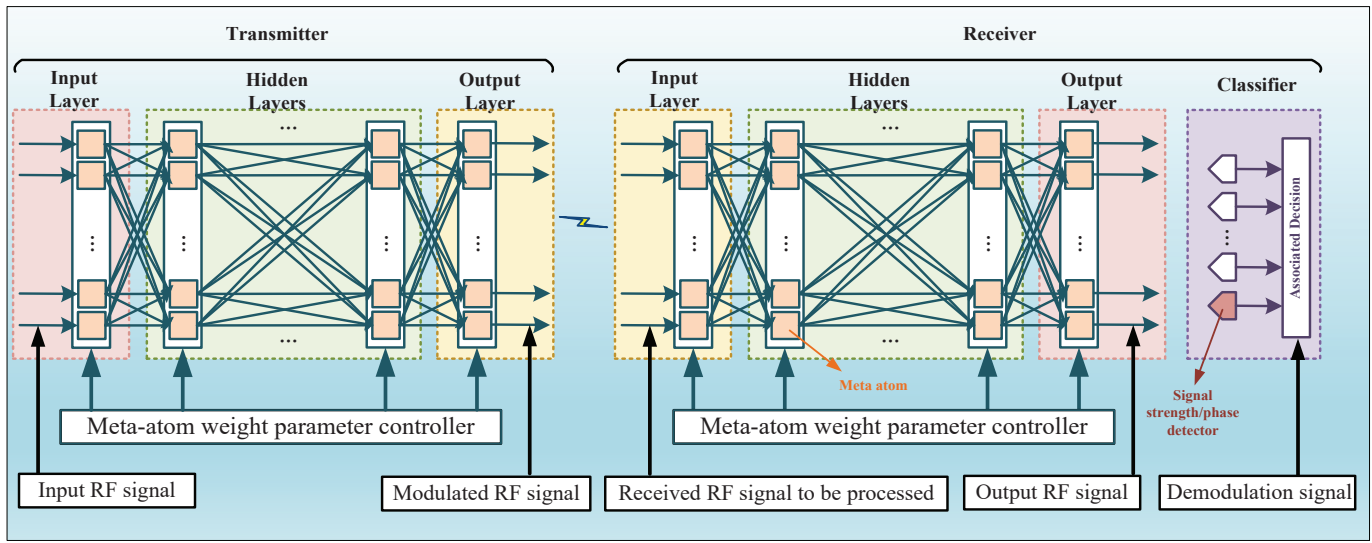}  
	\caption{Framework of the novel wireless transceiver based on Rb-DNN. }  
	\label{img9}  
	\vspace{-0.2cm}
\end{figure*} 

We can take a further step to explore the novel wireless transceiver on the basis of the RIS neural network structure, which can directly process the signal without RF transceiver modules.
Nowadays, conventional separate wireless transceiver architecture is widely used, including information processing module (baseband module) and RF transceiver module. 
However, conventional baseband modules cannot easily cope with the task of real-time processing of instantaneous massive baseband signals due to the extreme increase in baseband data throughput caused by the number of extreme large-scale MIMO and ultra-high bandwidth technologies. Moreover, the cost and power consumption of RF hardware also sharply increase. Therefore, the novel wireless transceiver, which integrates information processing and RF modules to achieve ultra-high-speed and low-power data transmission directly on RF signals, must be explored. 

In Fig.~\ref{img9}, the novel transceiver consists of an input layer, hidden layers, an output layer, and a classifier composed of a signal detector and an association decision unit. The classifier at the receiver first detects the spatial distribution of the intensity or phase of the output RF signal in the output layer, and then makes a joint decision. The classifier at the receiver first detects the spatial distribution of the intensity or phase of the output RF signal in the output layer and then makes an associated decision to obtain demodulated RF signals.
The RF signals are transmitted between the layers of the Rb-DNN. Each layer is composed of multiple meta-atoms. The meta-atom between the adjacent layers are interconnected via EM wave propagation. In specific transceiver tasks, the electromagnetic responses can be changed when the external control signal is loaded on the adjustable element of the meta atoms, and the weight parameters of the meta atom neurons are further reconfigured (i.e., the amplitude or phase change of EM wave transmission). Particularly, the processing speed of this type of transceiver may reach the speed of light and gains the advantages of high flexibility and low power consumption.

\subsection{Towards the intelligent sensing and localization system of RIS}
As previously mentioned, the intelligent manipulation of RISs is expected to achieve high-resolution sensing and high accuracy localization coexisting with the basic wireless communication functions, which can constantly share the abundant data and information in the time, frequency, and space domains.  Apart from providing extra reflecting paths to supplement LoS links in high frequency scenarios such as mmWave and THz ranges, RISs can enhance the power of NLoS links near the RIS deployment range. 

This study takes a low-energy reflective wall as an example. The power of the reflected links is particularly low to sense the channel information, which can be regarded as a blind area. Against the condition, acting as a highly reflective plane, an RIS can be deployed to enhance the energy of the reflected links for grabbing the channel information to expand the sensing range and enhance the coverage and further improve the localization.  Furthermore, AI and machine learning-aided can  extract deep features and hidden patterns of the raw data to increase the resistance to environmental noise doped in effective information and achieve high precision sensing and positioning which have not been seen before.  Not only for a single RIS, multiple RISs should be developed for further expanding the sensing scope. AI is a competitive method for the ultra-high deployment and computational complexities due to the multiple RISs by  jointly adjusting the configuration of each RIS to maximize the sensing range.

\subsection{Towards the RIS-assisted wireless communication system integrated with AI}

\subsubsection{Dynamic and flexible control of RIS}
Optimal performance can be obtained by dynamic beamforming of RIS. However, a key technical problem still persists, which is the manner by which, to dynamically and flexibly control the RIS. To the authors' knowledge, most studies on the RIS is based on the assumption that RISs are controlled by wired connection, which is relatively simple and low power consuming. In this case, wired connection routes need to be reserved and become the barrier for flexible deployment of RIS. Wireless connection between the BS and the controller avoids the constraints of wired connections. Nevertheless, the interface scheme may require additional protocols and increase the power consumption.

Under such a condition, we may consider a highly autonomous scheme in which the RIS controller independently adjusts according to the environment-aware information without the control of the server and interfaces. In combination with the sensing capability of sensors deployed in the surrounding RISs, the angle, direction, and even part of the channel information can be obtained, which may greatly reduce the difficulty of channel estimation and pilot overhead. Meanwhile, a large number of environmental characteristic information can be collected from environmental information to further enhance the training and deployment ability of AI and machine learning in wireless communication. However, the complete autonomous control of the RIS outside the control of the BS leads to difficulties in achieving joint estimation and beamforming with the BS, and the sensing capability and accuracy have a great effect on the system performance.

\subsubsection{AI data collection and model training}
Modern AI techniques have already provided various applications employed for the wireless transmission. Nevertheless, certain limitations remains to be solved. 
In these data-driven machine learning schemes, the effectiveness of the trained model heavily relies on the validity and quality of the data. The more accurate and abundant the data collected, the better the performance of the model gained. However, a large amount of labeled training data and high calculating power are required in those data-hungry supervised learning methods for training a well-established model, especially in the IWPE integrated with numerous RISs and sensors.
Therefore, semi-supervised learning, unsupervised learning, and label-free DRL could be stepped forward to learn from fast time-varying, reconfigurable environment, and applying the advances of generative adversarial networks to generate artificial data is a good approach.

Furthermore, the machine learning model is always time and power consuming, and it is trained offline before the online deployment. To overcome the cost of model training, the recent advances in domain adaptation and transfer learning are potential techniques for reducing the training workload of these models for optimization \cite{ali20206g}.

\section{Conclusion}
In this article, we comprehensively elaborate how RIS based communication frameworks and AI based technologies interact and promote with each other from several different aspects. Particularly, we discuss the recent studies on the novel RIS-based neural network architecture, intelligent imaging, recognition, sensing system of RIS based on ML methodology, and AI-embedded RIS-aided wireless communication systems. Several challenges are overcome by combining AI with RIS, 
which further make the RIS-controlled intelligent radio environments become a reality. However, significant additional research in RIS and AI are needed on a number of issues, including further exploration of RIS neural network-based wireless transceiver, RIS-based localization and sensing system design, dynamic and flexible control of  RIS, AI data collection and model training.

\bibliographystyle{IEEEtran}
\bibliography{reference}

\end{document}